\documentclass[12pt,a4paper]{article}

\usepackage{axodraw}
\usepackage{a4wide}
\usepackage{amsmath}
\usepackage{bm}
\usepackage{amssymb}
\usepackage{hyperref}
\usepackage{epic}

\newcommand{\cN}{{\cal N}}

\makeatletter
\def\mr@ignsp#1 {\ifx\:#1\@empty\else #1\expandafter\mr@ignsp\fi}%
\newcommand{\multiref}[1]{\begingroup
\xdef\mr@no@sparg{\expandafter\mr@ignsp#1 \: }%
\def\mr@comma{}%
\@for\mr@refs:=\mr@no@sparg\do{\mr@comma\def\mr@comma{,}\ref{\mr@refs}}%
\endgroup}
\makeatother

\begin{document}

\thispagestyle{empty}


\begin{center}
{\Large{\bf
The non-planar contribution\\[1mm]
to the four-loop universal anomalous dimension\\[3mm]
in $\cN=4$ Supersymmetric Yang-Mills theory
}}
\vspace{15mm}

{\sc
V.~N.~Velizhanin}\\[5mm]

{\it Theoretical Physics Department\\
Petersburg Nuclear Physics Institute\\
Orlova Roscha, Gatchina\\
188300 St.~Petersburg, Russia}\\[5mm]

\textbf{Abstract}\\[2mm]
\end{center}

\noindent{
We present the result of a \textit{full direct} component
calculation for the non-planar contribution to the four-loop anomalous dimension
of the Konishi operator in $\cN =4$ Supersymmetric Yang-Mills theory.
The result contains only $\zeta(5)$ term and proportional to $\zeta(5)$
contribution in the planar case, which comes purely from wrapping corrections.
We have extended also our previous calculations for the leading transcendental
contribution arXiv:0811.0607 on
non-planar case and have found the same results up to a common factor. It
allows us to suggest that the non-planar contribution to the four-loop universal
anomalous dimension for the twist-2 operators with arbitrary Lorentz spin
is proportional to $S_1^2(j)\, \zeta(5)$.
This result gives unusual double-logarithmic asymptotic $\ln^2\!j$ for large $j$.
}
\newpage

\setcounter{page}{1}


In our previous papers~\cite{Velizhanin:2008jd,Velizhanin:2008pc}
 we have calculated the planar four-loop anomalous
dimension of the Konishi operator and the leading transcendental contribution
to the four-loop universal anomalous dimension of twist-2 operators in the $\cN=4$ supersymmetric
Yang-Mills (SYM) theory. Calculations have been performed in component as a {\it full direct}
computation of the anomalous dimension of the twist-2 operator.
The advantages of our method are the full automation of process of calculations and
the absence of any suggestion about specific properties of operators, so these
results can serve as ``experimental'' test
for the similar results obtained with the help of integrability~\cite{Minahan:2002ve} in the
framework of AdS/CFT-correspondence~\cite{Maldacena:1997re}.
The planar four-loop anomalous dimension of the Konishi operator were computed
earlier by two
different ways from the both sides of AdS/CFT-correspondence. In the $\cN=4$ SYM
theory the calculations were performed in the superfield formalism and take
into account only diagrams did not included in the asymptotic Bethe-ansatz~\cite{Fiamberti:2007rj},
following Ref.~\cite{Sieg:2005kd}.
From superstring side~\cite{Bajnok:2008bm} the finite size effects
were take into account
using L\"{u}scher formulas~\cite{Luscher:1986pf}.
The results of both computations are in agreement after corrections from the
perturbative side.  Our result of the {\it full direct} calculation is the same and
confirms correctness all suggestions of the computations from
Refs.~\cite{Fiamberti:2007rj,Bajnok:2008bm}, including the correctness of the
asymptotic Bethe-ansatz up to the four loops.
Our result for leading transcendental contribution to the four-loop universal
anomalous dimension for the twist-2 operators with arbitrary Lorentz spin was used
together with the predictions from the Balitsky-Fadin-Kuraev-Lipatov (BFKL) equation~\cite{Lipatov:1976zz} to check
the corresponding part of the finite size corrections~\cite{Bajnok:2008qj} to the four-loop
anomalous dimension of BMN-operators~\cite{Berenstein:2002jq} from the $sl(2)$ sector
obtained from the asymptotic Bethe ansatz~\cite{Kotikov:2007cy}.

One more advantage of our method of the {\it full direct} calculation
is the possibility to
obtain full color structure for the four-loop Konishi. In this paper we present from the first
time the result for the non-planar (color subleading) contribution to the four-loop Konishi and
for the first three even moments for the leading transcendental contribution to the four-loop
universal anomalous dimension of the twist-2 operators with arbitrary Lorentz spin in $\cN=4$ SYM theory.

The calculation of the non-planar contribution, which is proportional to the quartic
Casimir operator $d_{44}$ (see Ref.~\cite{vanRitbergen:1998pn}), can be split in
two part according to the basic parent topology for master-integrals.
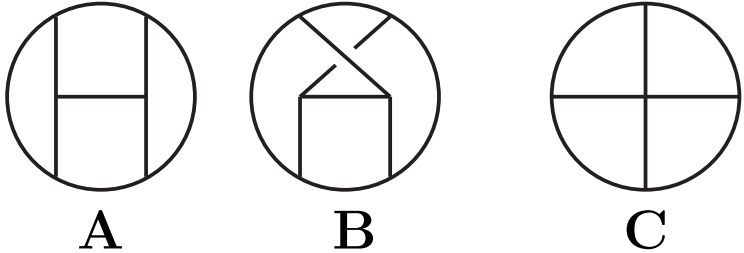
\begin{figure}[ht]
\vspace{-1.5cm}
\begin{center}
\begin{picture}(700,150)(0,0)
\SetScale{0.7}
\SetWidth{2}
\Line(26,7)(26,93)
\Line(74,7)(74,93)
\Line(26,50)(74,50)
\CArc(50,50)(50,0,360)
\Text(35,-15)[c]{{\LARGE {\bf A}}}
\Line(156,7)(156,50)
\Line(156,50)(175,67)
\Line(185,76)(204,93)
\Line(204,7)(204,50)
\Line(204,50)(156,93)
\Line(156,50)(204,50)
\CArc(180,50)(50,0,360)
\Text(130,-15)[c]{{\LARGE {\bf B}}}
\Line(340,0)(340,100)
\Line(290,50)(390,50)
\CArc(340,50)(50,0,360)
\Text(239,-15)[c]{{\LARGE {\bf C}}}
\end{picture}
\end{center}
\vspace{10mm}
\caption{Basic parent four-loop planar ({\bf A}) and non-planar ({\bf B}) topologies and
  topology~({\bf C}) obtained by cancelling one
  horizontal line in topology {\bf A} or any
  one line in topology~{\bf B}.}
\label{Toplogy}
\end{figure}
For the
four-loop tadpoles with equal mass lines we have two parent topologies
presented in Fig.~\ref{Toplogy}: planar topology {\bf A} and non-planar topology {\bf B}. These two
topologies correspond to the master-integrals PR12 and PR0 from Table 1 of Ref.~\cite{Czakon:2004bu}
and all other topologies from this table can be obtained by cancelling one and more
lines inside parent topologies. For the diagrams from the first, planar class
we already have the well-tested program for the four-loop calculations
which based on our own implementation of the Laporta's algorithm~\cite{Laporta:2001dd} for the resolution
of the integration by part (IBP) identities in
the form of MATHEMATICA package BAMBA and using the method from Refs.~\cite{Misiak:1994zw}.
With this program we produced database
for all necessary integrals expressed through master-integrals from
Ref.~\cite{Czakon:2004bu}.
For the second, non-planar class of diagrams in principal we should generate and
resolve all sets of new IBP identities.
But we have a great simplification if noting that the cancellation any one line
in the non-planar topology {\bf B} gives topology {\bf C}, which can be obtained with the cancellation of
any horizontal line from the planar topology {\bf A}
and for topology {\bf C} we already have the recurrence relations for all necessary integrals.
So, for the non-planar topology {\bf B} we should go only on one step in the resolution of the IBP identities,
then change momentum according to topology {\bf C} and reexpand denominators if it is necessary.
In this way we extend our database only with integrals with different positive
powers for all nine propagators of the non-planar topology {\bf B}.

Thus, we have repeated our previous calculations for the Konishi operator~\cite{Konishi:1983hf}
\begin{equation}
\mathcal{O}_\mathcal{K}=\mathrm{tr}\left[A_{i}A^{i}+B_{i}B^{i}\right],\qquad i=1,2,3\,,
\end{equation}
where
$A^i$ and $B^i$ are the real adjoint scalar and pseudoscalar fields correspondingly,
keep quartic Casimir operator $d_{44}$ (did not substitute its leading color value
$d_{44}=N^2_c/24$ as before) and using additional FORM procedure for the diagrams with the
non-planar topology {\bf B} from Fig.~\ref{Toplogy}. To check, that our program work
correctly we reproduce both planar ($\sim C_A^4$) and non-planar ($\sim
d_{44}$) parts for the anomalous dimension of the gluon
field~\cite{Czakon:2004bu}
coming from the pure Yang-Mills gauge theory.
All diagrams were produced with DIANA~\cite{Tentyukov:1999is} , which
call QGRAF~\cite{Nogueira:1991ex} to generate all diagrams and obtained code were
calculated with FORM~\cite{Vermaseren:2000nd},
using FORM package COLOR~\cite{vanRitbergen:1998pn} for evaluation of the color traces.

The final result
after subtraction of the anomalous dimension for the scalar fields is:
\begin{equation}
\gamma^{\mathrm{4-loop}}_{\mathcal{K}}
\ = \
12\,g^2
-48\,g^4+336\,g^6+\left(-2496+576\,\zeta(3)-1440\left(1+\frac{12}{ N_c^2}\right)\zeta(5)\right)\,g^8\,,\label{resk}
\end{equation}
with
\begin{equation}
g^2=\frac{g^2_{YM}N_c}{(4\pi)^2}\,,\qquad d_{44}=\frac{N_c^2(N_c^2+36)}{24}
\end{equation}
and the following non-planar contribution to the four-loop anomalous dimension for the (pseudo)scalar fields
(see Ref.~\cite{Velizhanin:2008jd} for the planar part):
\begin{equation}
\gamma^{\mathrm{4-loop,}\;{\mathrm{non-planar}}}_{\phi}=\Big(-42 - 177\, \zeta(3) + 555\, \zeta(5)\Big)\frac{g^8}{N_c^2}\,.
\end{equation}
An important check of our result (\ref{resk}) is the absence of higher poles
and some special numbers such as $\zeta(2)$, $\zeta(4)$, $S_2$ and other, which enter in the scalar master integrals
from Ref.~\cite{Czakon:2004bu}.

Also we have repeated our previous calculations for the first three moments of the leading
transcendental contribution to the four-loop universal anomalous dimension of twist-2 operators~\cite{Velizhanin:2008pc}.
As in planar case we have produced the database
for the scalar integrals\footnote{Results for integrals can be obtained under request.}
containing $\zeta_5$ for non-planar topology {\bf B} from Fig.~\ref{Toplogy}
with the MATHEMATICA package FIRE~\cite{Smirnov:2008iw}.
Surprisingly, but the non-planar leading transcendental contribution to the four-loop
universal anomalous dimension is modified in the same way as the Konishi in Eq.~(\ref{resk})
\begin{equation}\label{HSZ5ResS1}
\hat\gamma_{uni}^{(3)}(j)\ =\ -\,640\,S_{1}^2(j-2)\left(1+\frac{12}{N_c^2}\right)\zeta_5\,.
\end{equation}

Based on the fact, that the non-planar contribution to the four-loop anomalous dimension of the Konishi
operator~(\ref{resk}) contains only $\zeta(5)$ and take into account Eq.~(\ref{HSZ5ResS1})
it is reasonable to suggest that the {\it full}
non-planar (color subleading) contribution to the four-loop universal
anomalous dimension of twist-2 operators with arbitrary Lorentz spin has the following
form:
\begin{eqnarray}
\gamma^{(3)}_{uni,\;np}\,(j)&\ =&\
-640\; S_1^2(j-2)\,\frac{12}{N_c^2}\ \zeta(5)\,,
\label{resnpuad}
\end{eqnarray}
with\footnote{Similar result was obtained for the twist-3 operator up to the
five loops~\cite{Kotikov:2007cy,Beccaria:2007cn}.}
\begin{equation}
\gamma_{uni}(j)=\gamma_{uni}^{(0)}(j)\;g^2
+\gamma_{uni}^{(1)}(j)\;g^4
+\gamma_{uni}^{(2)}(j)\;g^6
+\gamma_{uni}^{(3)}(j)\;g^8
+...
\end{equation}
from Refs.~\cite{LN4,Kotikov:2003fb,Kotikov:2004er,Kotikov:2007cy,Bajnok:2008qj}.

Note, that really, we have calculated only first three even moments $j=2$, $j=4$ and $j=6$
for this equation and it can be modified with other harmonic sums (see Ref.~\cite{Velizhanin:2008pc}).
Moreover, it is possible, that for higher values of moments $j$
the non-planar contribution to the four-loop universal anomalous dimension will
contain
the terms which is proportional to $\zeta(3)$ and rational number to
give such combination of the harmonic sums that for $j=4$, i.e. for the Konishi, they will give zero.

However, let's us suggest that Eq.(\ref{resnpuad}) is correct. There are two important
consequence of this result. First of all it will gives the unusual behavior for the
large spin $j$ limit, which will proportional to $\ln^2\!j$ instead of
expected $\ln j$~\cite{Korchemsky:1988si}.
Another consequence comes from the analytical continuation to $j\to 1+\omega$,
where the predictions from the BFKL equation exist. The analytical continuation of Eq.(\ref{resnpuad}) gives
$1/\omega^2$ pole which comes from the (jet unknown) next-to-next-to-leading logarithmic
approximation (NNLLA or third order corrections) to the
kernel of the BFKL equation and, then, it should contain corresponding non-planar
contribution at this order.
As the universal anomalous dimension of the twist-2 operators and the kernel of the BFKL equation in $\cN=4$ SYM is the
most complicated part of the corresponding QCD results these futures should be hold also in QCD.

In spite of planar AdS/CFT system well studied from the both sides of
gauge/string duality and seems to be solved~\cite{Gromov:2009tv},
non-planar results in $\cN=8$ SYM theory have been obtained only for
gluon scattering amplitudes, which
are related with $\cN=8$ Supergravity~\cite{Green:1982sw}.
Our results show that the non-planar contribution at least for the anomalous dimension of the twist-2
operators has a rather simple form and
relation
with the wrapping corrections
and probably can be studied with other methods widely used
for the planar case.

 \subsection*{Acknowledgments}
We would like to thank Lev Lipatov, Andrei Onishchenko and Matthias Staudacher for useful discussions.
This work is supported by
RFBR grants 07-02-00902-a, RSGSS-3628.2008.2.
We thank the Max Planck Institute for Gravitational Physics
at Potsdam for hospitality while working on parts of this
project.

\end{document}